\documentclass[11pt]{article}
\usepackage{graphicx,ctable}
\usepackage{amssymb}
\usepackage{hyperref}
\usepackage{float}
\usepackage{multirow}
\newcommand {\dNchdeta} {\ensuremath{\mathrm{d}N_\mathrm{ch}/\mathrm{d}\eta }}
\newcommand {\Tch} {\ensuremath{T_\mathrm{ch}}}
\begin{document}
\begin{center}
{\Large\bf{Thermodynamic limit in high-multiplicity proton-proton collisions \\ at $\sqrt{s}$ = 7 TeV}}\\
\vspace*{1cm}  
{\sf{Natasha Sharma$^1${\footnote{email: natasha.sharma@cern.ch}}, Jean Cleymans$^2$ and
Boris Hippolyte$^3$}}\\[0.1cm]
\vspace{10pt}  
{\small  {\em 
$^1$ Department of Physics, Panjab University, Chandigarh 160014, India\\
$^2$UCT-CERN Research Centre and Department of Physics, University of Cape Town,\\ Rondebosch 7701, South Africa\\
$^3$Institut Pluridisciplinaire Hubert Curien and Universit\'e de Strasbourg Institute for Advanced Study, CNRS-IN2P3, Strasbourg, France}}
\normalsize  
\end{center}  

\begin{abstract}
An analysis is made of the particle composition in the final state of proton-proton (pp) collisions
at 7~TeV as a function of the charged particle multiplicity ($\dNchdeta$). The thermal model is used
to determine the chemical freeze-out temperature as well as the radius and strangeness suppression
factor $\gamma_s$. Three different ensembles are used in the analysis. The grand canonical ensemble,
the canonical ensemble with exact strangeness conservation and the canonical ensemble with exact
baryon number, strangeness and electric charge conservation. It is shown that for the highest
multiplicity class the three ensembles  lead to the same result. This allows us to conclude that this
multiplicity class is close to the thermodynamic limit. It is estimated that the final state in pp collisions
could reach the thermodynamic limit when  $\dNchdeta$ is larger than twenty per unit of rapidity,
corresponding to about 300 particles in the final state when integrated over the full rapidity interval.
\end{abstract}
%
\section{\label{secIntroduction}Introduction}
%
In statistical mechanics the thermodynamic  limit is  the limit in which the total number of particles $N$
and the volume $V$ become large but the ratio $N/V$ remains finite and results obtained in the microcanonical, 
canonical and  grand canonical ensembles become equivalent. In this paper we argue that
this limit might be reached in high energy pp collisions if the total number of charged hadrons becomes
larger than 20 per unit of rapidity in the mid-rapidity region, corresponding to roughly 300 particles
in the final state when integrated over the full rapidity interval. 
For this purpose use is made of the data published by the ALICE Collaboration~\cite{ALICE:2017jyt} on the
production of multi-strange hadrons in pp collisions as a function of charged particle multiplicity in a one unit
pseudorapidity interval $\langle \dNchdeta \rangle|_{|\eta|<0.5}$.
These data have attracted significant attention because they cannot be reproduced by standard Monte Carlo 
models~\cite{Sjostrand:2007gs,Pierog:2013ria,Bierlich:2015rha}.\\

In high energy collisions applications of the statistical model in the form of the hadron resonance gas
model have been successful~\cite{Andronic:2017pug,Becattini:2017pxe} in describing the composition of the final state e.g. the yields of pions,
kaons, protons and other hadrons. In these descriptions use is made of the grand canonical ensemble
and the canonical ensemble with exact strangeness conservation. In this paper we consider in addition
the use of the canonical ensemble with exact baryon, strangeness and charge conservation.

The identifying feature of the thermal model  is that all the resonances  listed in~\cite{Patrignani:2016xqp}  are
assumed to be in thermal and chemical equilibrium.
This  assumption drastically reduces the number of free parameters as this stage is determined by just a few
thermodynamic variables namely, the chemical freeze-out temperature $T_{ch}$, the various chemical potentials $\mu$ determined by
the conserved quantum numbers and by the volume $V$ of the system.
It has been shown that this description is also the correct
one~\cite{Cleymans:1997eq,Broniowski:2001we,Akkelin:2001wv} for a scaling expansion as first discussed by
Bjorken~\cite{Bjorken:1982qr}.  
After integration over $p_T$ these authors have shown that:
\begin{equation}
{dN_i/dy\over dN_j/dy} = {N_i^0\over N_j^0} 
\end{equation}
where $N_i^0$ is the particle yield
 as calculated in a fireball at rest.
Hence, in the Bjorken model with longitudinal scaling and radial expansion the effects of hydrodynamic flow cancel out in ratios.

We will show in this paper that the difference between the ensembles used disappears if the final state multiplicity is large. All calculations 
were done using THERMUS~\cite{Wheaton:2004qb}.

We compare three different ensembles based on the thermal model.
\begin{itemize}
\item Grand canonical ensemble (GCE), the conservation of quantum numbers is implemented using chemical potentials. 
The quantum numbers are conserved on the average.
The partition function depends on thermodynamic quantities and the Hamiltonian describing the system of $N$ hadrons: 
\begin{equation}
Z_{GCE} = \textrm{Tr} \left[ e^{-(H-\mu N)/T}\right]
\end{equation}
which, in the framework of the thermal model considered here, leads to
\begin{equation}
\ln Z_{GCE}(T,\mu,V) = \sum_i g_i V \int\frac{d^3p}{(2\pi)^3}\exp\left( -\frac{E_i-\mu_i}{T} \right)\\
\end{equation}
in the Boltzmann approximation.
The yield is given by:
\begin{equation}
N_i^{GCE} = V \int \frac{d^3p}{(2\pi)^3} \exp \left( -\frac{E_i}{T}\right)  .
\end{equation}
We have put the chemical potentials equal to zero, as relevant for the beam energies considered here.
The decays of resonances have to be added to the final yield
\begin{equation}
N_i^{GCE}(\mathrm{total}) = N_i^{GCE} + \sum_j Br(j\rightarrow i) N_i^{GCE}  .
\end{equation}

\item Canonical ensemble with exact implementation of strangeness conservation, we will 
refer to this as the strangeness canonical ensemble (SCE). 
There are chemical potentials for baryon number $B$ and charge $Q$ but not for strangeness:
\begin{equation}
Z_{SCE} = \textrm{Tr}\left[ e^{-(H-\mu N)/T}\delta_{(S,\sum_iS_i)}\right]
\end{equation}
The delta function imposes exact strangeness conservation, requiring overall strangeness to be fixed to the value $S$, in this paper
we will only consider the case where overall strangeness is zero, $S=0$.
This change leads to~\cite{BraunMunzinger:2001as}:
\begin{equation}
Z_{SCE} = \frac{1}{(2\pi)}
\int_0^{2\pi} d\phi e^{-iS\phi}
Z_{GCE}(T,\mu_B,\lambda_S)
\end{equation}
where the fugacity factor is replaced by
\begin{equation}
\lambda_S = e^{i\phi}
\end{equation}
\begin{eqnarray}
N_{i}^{SCE}&=&V{{Z^1_{i}}\over {Z_{S=0}^C}} \sum_{k,p=-\infty}^{\infty} a_{3}^{p} a_{2}^{k}
a_{1}^{{-2k-3p- s}} I_k(x_2) I_p(x_3) I_{-2k-3p- s}(x_1),   \label{equ6}
\end{eqnarray}
where $Z^C_{S=0}$ is the canonical partition function
\begin{eqnarray}
Z^C_{S=0}&=&e^{S_0} \sum_{k,p=-\infty}^{\infty} a_{3}^{p}
a_{2}^{k} a_{1}^{{-2k-3p}}\nonumber 
 I_k(x_2) I_p(x_3) I_{-2k-3p}(x_1),
\label{eq7}
\end{eqnarray}
where $Z^1_i$ is the one-particle partition function calculated for $\mu_S=0$ in the Boltzmann
approximation. The arguments of the Bessel functions $I_s(x)$ and the parameters $a_i$ are introduced as,
\begin{eqnarray} a_s= \sqrt{{S_s}/{S_{\mathrm{-s}}}}~~,~~ x_s = 2V\sqrt{S_sS_{\mathrm{-s}}} \label{eq8a}, \end{eqnarray}
where $S_s$ is the sum  of all $Z^1_k(\mu_S=0)$  for particle species $k$ carrying strangeness
$s$. 
As previously, the decays of resonances have to be added to the final yield
\begin{equation}
N_i^{SCE}(\mathrm{total}) = N_i^{SCE} + \sum_j Br(j\rightarrow i) N_i^{SCE}  .
\end{equation}

\item Canonical ensemble with exact implementation of $B$, $S$ and $Q$ conservation, we will refer to this as the 
full canonical ensemble (FCE). 
In this ensemble 
there are no chemical potentials. The partition function is given by:
\begin{equation}
Z_{FCE} = \textrm{Tr}\left[ e^{-(H-\mu N)/T}\delta_{(B,\sum_iB_i)}\delta_{(Q,\sum_iQ_i)}\delta_{(S,\sum_iS_i)}\right]
\end{equation}
\begin{equation}
Z_{FCE} = \frac{1}{(2\pi)^3}
\int_0^{2\pi} d\psi e^{-iB\alpha}
\int_0^{2\pi} d\phi e^{-iQ\psi}
\int_0^{2\pi} d\alpha e^{-iS\phi}
Z_{GCE}(T,\lambda_B,\lambda_Q,\lambda_S)
\end{equation}
where the fugacity factors have been replaced by
\begin{equation}
\lambda_B = e^{i\alpha},\quad \lambda_Q = e^{i\psi}, \quad \lambda_S = e^{i\phi}   .
\end{equation}
As before, the decays of resonances have to be added to the final yield
\begin{equation}
N_i^{FCE}(\mathrm{total}) = N_i^{FCE} + \sum_j Br(j\rightarrow i) N_i^{FCE}  .
\end{equation}

A similar analysis was done in~\cite{Abelev:2006cs} for pp collisions at 200 GeV but without the  dependence 
 on charged multiplicity.
\end{itemize}
In this case the analytic expression becomes very lengthy and we refrain from writing it down here, it is implemented in the 
THERMUS program~\cite{Wheaton:2004qb}.

These three ensembles are applied to pp collisions in the central region of rapidity. 
It is well known that in this kinematic region, 
one has particle - antiparticle symmetry and therefore there is no net baryon density and also no net strangeness. The different ensembles nevertheless
give different results because of the way they are implemented. A clear size dependence is present in the results of the ensembles.
In the thermodynamic limit they should become equivalent. 
Clearly there are other ensembles that could be investigated and also other sources of finite volume corrections. We hope to address these
in a longer publication in the near future. 

A similar analysis was done in~\cite{Abelev:2006cs,Becattini:2009ee,Becattini:2010sk} for pp collisions at 200 GeV but 
without the  dependence 
 on charged multiplicity.


\section{Comparison of different statistical ensembles.}
\begin{figure}
\begin{center}
\includegraphics[width=0.7\textwidth,height=16cm]{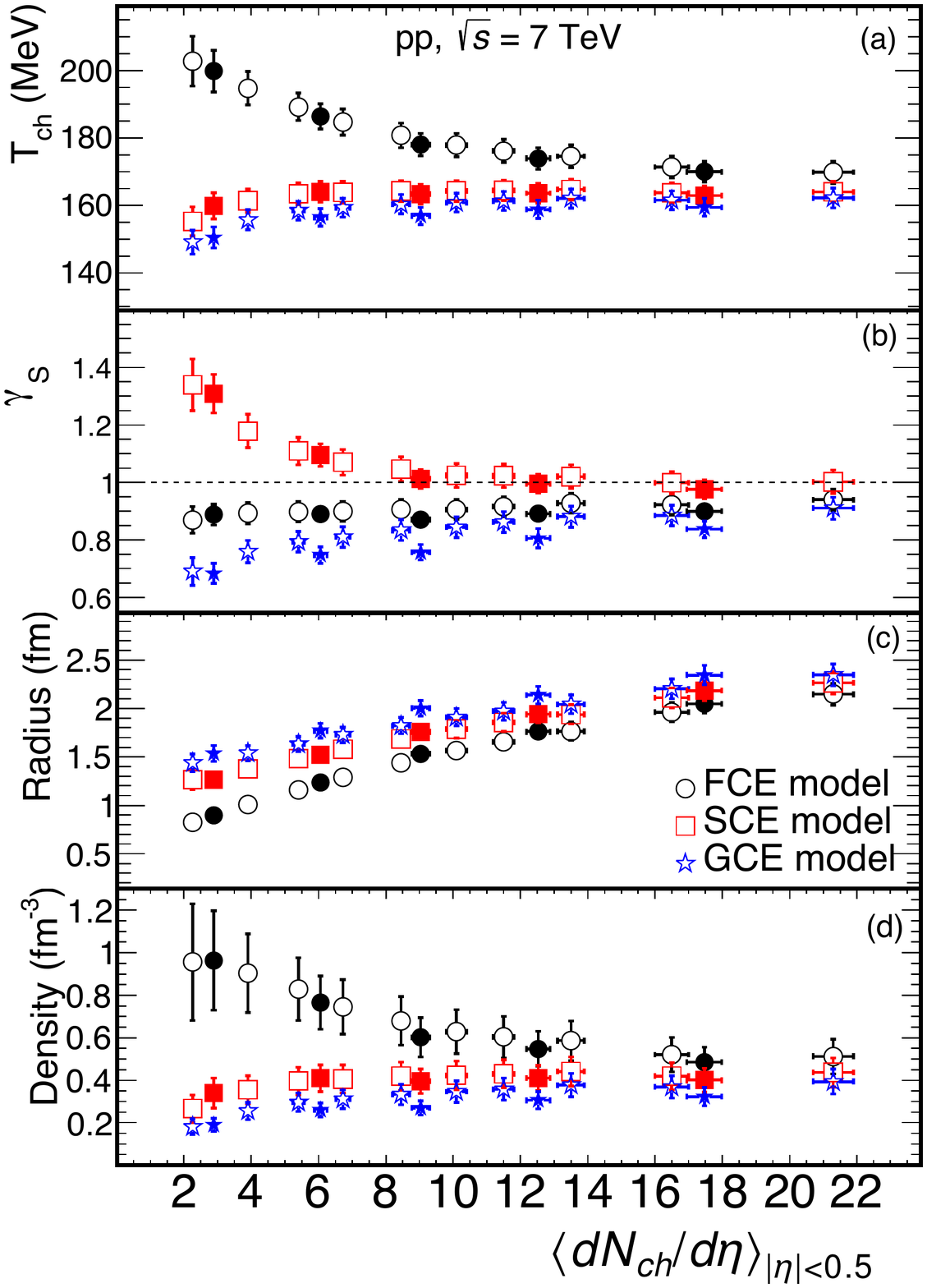}
\caption{The chemical freeze-out temperature $\Tch$ obtained for three different ensembles in the upper panel (a).  
The strangeness suppression
factor, $\gamma_s$ is shown in panel (b). The radius of the system at chemical freeze-out is shown in panel (c). 
The density is shown in the bottom panel (d). The open symbols show results of fitting hadrons yields without $\Omega$ whereas solid symbols 
show fit results including $\Omega$ yields.
}
\label{T_pp}
\end{center}
\end{figure}

%
%
In Fig.~\ref{T_pp}a we show the chemical freeze-out temperature as a function of the multiplicity of hadrons in the final state~\cite{ALICE:2017jyt}.
The freeze-out temperature has been calculated using three different ensembles. 
The highest values are obtained using the canonical ensemble
with exact conservation of three quantum numbers, baryon number $B$, strangeness $S$ and charge $Q$, all of them
being set to zero as is appropriate for the central rapidity region in pp collisions at 7~TeV. 
In this ensemble the temperature drops 
very clearly from the lowest to the highest multiplicity intervals. 
The open symbols in Fig.~\ref{T_pp} were calculated using as input the 
yields for $\pi^+ + \pi^-$, $\mathrm{p} + \bar{\mathrm{p}}$, $K^0_S$, $\Lambda + \bar{\Lambda}$ and $\Xi^- + \bar{\Xi}^+$ while the full symbols also 
include the yields for $\Omega^- + \bar{\Omega}^+$ as given in~\cite{ALICE:2017jyt}~\footnote{The values used in this study were obtained by the ALICE Collaboration and can be
found at the url: \href{https://www.hepdata.net/record/77284}{https://www.hepdata.net/record/77284}.}.
As an example we show a comparison between measured and fitted values for the multiplicity class II in table 2.  

\begin{table}[H]
  \centering
	\caption{Comparison between measured and fitted values for pp collisions at 7 TeV for V0M multiplicity class II.} 
	\label{tab:cent-event}\vspace{0.1in}
	\begin{tabular}{|c|c|c|c|c|}
	\hline	
	\multirow{2}{*}{Particle Species} & \multirow{2}{*}{$dN/dy$ (data)} & \multicolumn{3}{c|}{$dN/dy$ (model)} \\
	\cline{3-5}
	&       &  Canonical S  &  Canonical B, S, Q & Grand Canonical  \\
		\hline
	$\pi^+$ & 7.88   $\pm$ 0.38   &  6.78 & 6.76 &   6.96 \\	
	\hline
	$K^0_S$ &  1.04    $\pm$ 0.05 & 1.16 & 1.16 & 1.15  \\
	\hline
	 $p$ &  0.44     $\pm$ 0.03  & 0.50 & 0.50 & 0.50\\
	\hline
	$\Lambda$ & 0.302     $\pm$  0.020 & 0.259 & 0.262 &  0.246 \\
	\hline
	$\Xi^{-}$ & 0.0358    $\pm$  0.0023  & 0.035 & 0.035  &  0.036 \\
	\hline
	\end{tabular}
\end{table}

The lowest values for $\Tch$ are obtained when using the grand canonical ensemble, in this case the conserved quantum numbers are 
again zero but only in an average sense.
The results are clearly different from those obtained in the previous ensemble, especially in the low multiplicity intervals. They gradually approach
each other and they become equivalent at the highest multiplicities.

For comparison with the previous two cases we also calculated $\Tch$ using the canonical ensemble with only
strangeness $S$ being exactly conserved using the method presented in~\cite{BraunMunzinger:2001as}. In this case the results are very close to those 
obtained in the grand canonical ensemble, with the values of $\Tch$ always slightly higher than in the grand canonical ensemble. Again for the 
highest multiplicty interval the results become equivalent.
As can be seen in the upper panel, Fig.~\ref{T_pp}a, even though all the ensembles produce different results, for high multiplicities the results 
converge to a common value around 160 MeV.

In Fig.~\ref{T_pp}b  we show results for the strangeness suppression factor $\gamma_s$ first introduced in~\cite{Letessier:1993hi}. 
In this case we obtain again quite substantial differences in each
one of the three ensembles considered. The highest values being found in the canonical ensemble with exact strangeness conservation.
Note that the values of $\gamma_s$ converge to unity as common value, i.e. full chemical equilibrium. 

In Fig.~\ref{T_pp}c the radius at chemical  freeze-out obtained in the three ensembles is presented. As in the previous figures, the results become 
independent of the ensemble chosen for the highest multiplicities.

An interesting feature is that the volume at chemical freeze-out increases linearly with the multiplicity in the final state.
This means that the density at chemical freeze-out tends to a  constant for high multiplicities. 
Again the three ensembles tend to a common value for the highest multiplicity class.
This is shown in the bottom panel, Fig.~\ref{T_pp}d where the ratio $(\dNchdeta)/\frac{4\pi R^3}{3}$  of the system at 
chemical freeze-out is plotted. 

The results in Fig.~\ref{T_pp} show that there is a strong correlation between some of the parameters. The very high temperature obtained in the canonical $BSQ$
ensemble correlates with the small radius in the same ensemble. Particle yields increase with temperature but a small volume
decreases them, hence the correlation between the two parameters.

\begin{figure}
\begin{center}
\includegraphics[width=0.7\textwidth,height=16cm]{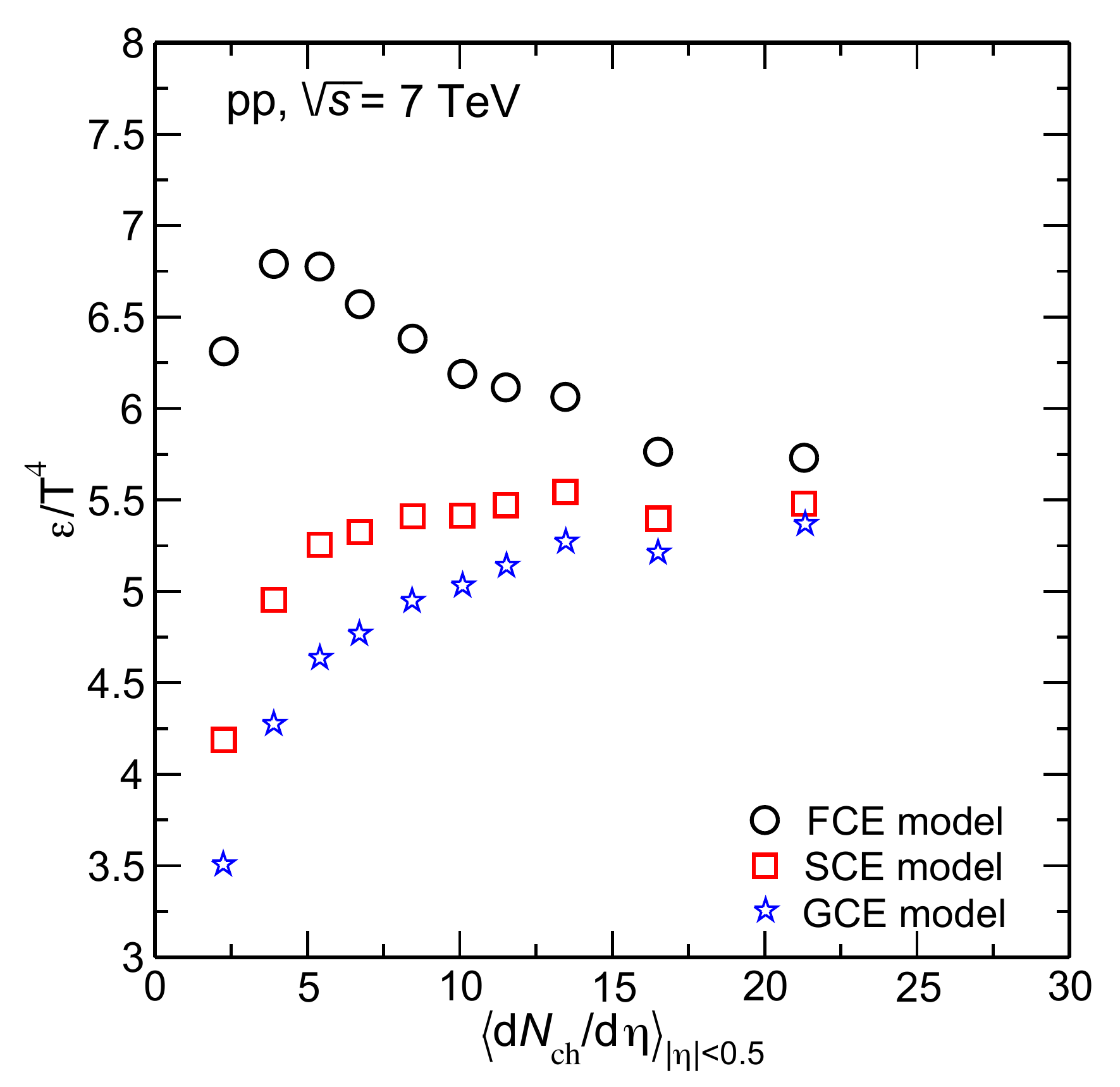}
\caption{The energy density $\varepsilon/T^4$ obtained for three different ensembles.  
}
\label{energy_pp}
\end{center}
\end{figure}

For completeness we also calculated  the energy density $\varepsilon /T^4$ using the three ensembles as this plays a role in
many theoretical considerations.  The results obtained are 
shown in Fig.~\ref{energy_pp} and are in line with those in Fig.~\ref{T_pp} for the particle density with a convergence
to the same energy density for the three different ensembles at the highest multiplicities.

\begin{figure}
\begin{center}
\includegraphics[width=0.8\textwidth,height=16cm]{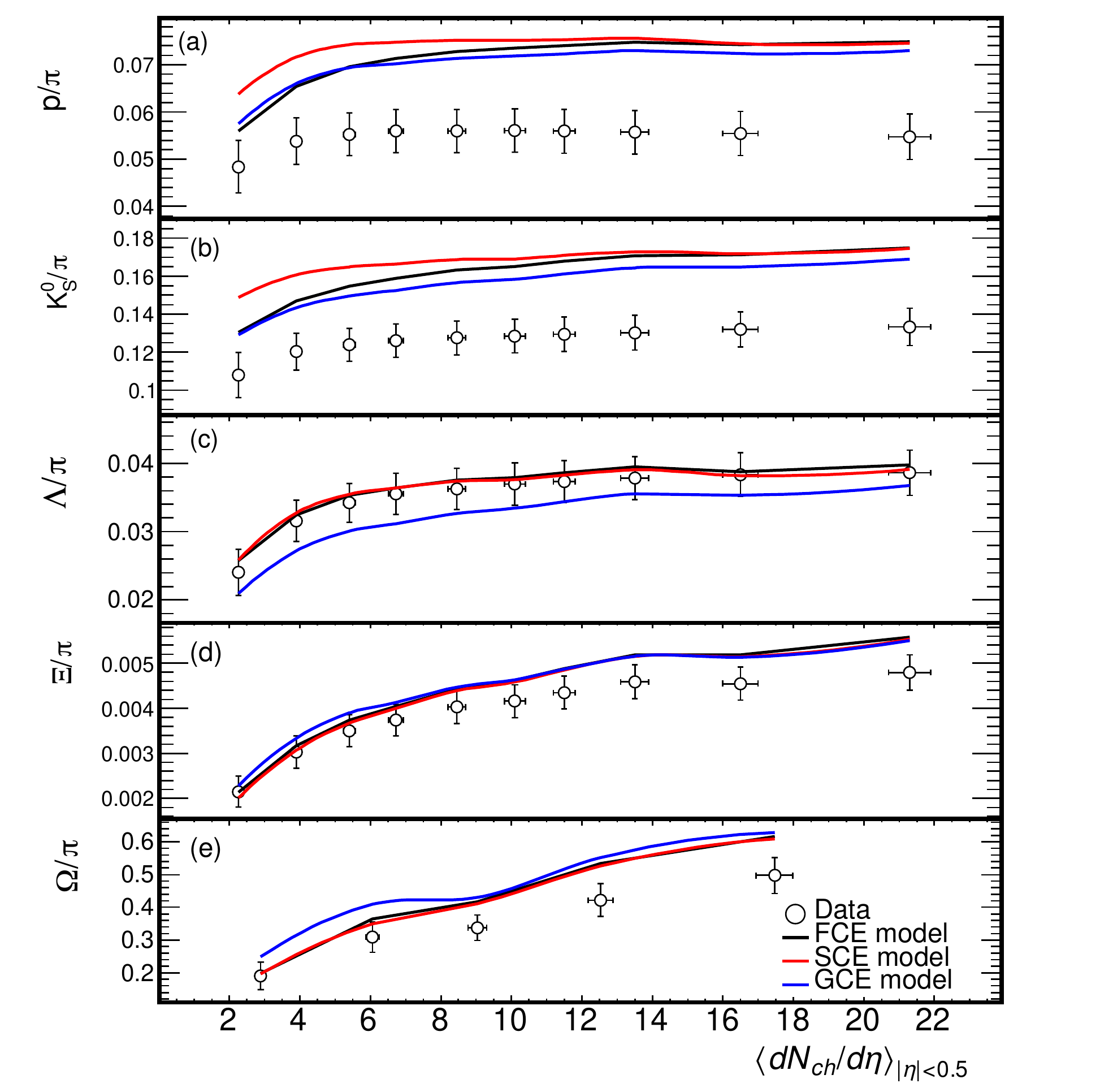}
\caption{
Ratios of particle to pion yields as a function of the final-state multiplicity.  
}
\label{proton2pion}
\end{center}
\end{figure}

In Fig.~\ref{proton2pion} we show the ratios of particle yields to the pion yields for three different ensembles. Deviations are caused by the
known underestimation of pion yield in the thermal models. 
The comparison of $\Omega/\pi$ ratio data with three different ensembles is shown in  Fig.~\ref{proton2pion}e for the case when $\Omega$ is included in the fits.
%
%
%
%
%

Table 2 shows the $\chi^2$ values obtained for the three ensembles considered in this paper.
\begin{table}[ht]
\begin{center}
\begin{tabular}{|c|c|c|c|}
\hline
$\langle dN_{ch}/d\eta \rangle|_{|\eta|<0.5}$& Canonical S  & Canonical B, S, Q & Grand Canonical      \\
\hline 
2.89   	 &  	6.04 / 3	&       24.29 / 3	&       29.05 / 3 \\
6.06	      	 &	16.02 / 3	&       25.89 / 3	&       32.28 / 3 \\
9.039	  & 	21.53 / 3	&       25.44 / 3	&       34.58 / 3 \\
12.53	   &	23.83 / 3	&       25.08 / 3	&       27.45 / 3 \\
17.47	   &	23.73 / 3	&       15.93 / 3	&       11.81 / 3 \\
\hline
\hline

2.26  &   3.85 / 2 & 12.79 / 2  & 6.45 / 2 \\
3.9   &   9.15 / 2 & 20.16 / 2 & 14.47 / 2 \\
5.4   &   14.94 / 2 & 25.46 / 2 & 20.27 / 2 \\
6.72  &   16.58 / 2 & 24.61 / 2 & 20.09 / 2 \\
8.45  &   18.71 / 2 & 24.65 / 2 & 20.83 / 2 \\
10.08 &   20.03 / 2 & 24.45 / 2 & 21.61 / 2 \\
11.51 &   20.91 / 2 & 24.42 / 2 & 21.80 / 2 \\
13.46 &   22.25 / 2 & 24.84 / 2 & 22.46 / 2 \\
16.51 &   22.19 / 2 & 23.52 / 2 & 22.41 / 2 \\
21.29 &   21.83 / 2 & 22.20 / 2 & 21.55 / 2 \\

\hline  
\end{tabular}
\caption{Values of $\chi^2$/ndf for various fits. The values in the top (bottom) part include (exclude) the $\Omega$ yields in the fits.
}
\end{center}
\end{table}
%
%
\section{Discussion and Conclusions}
%
%
%
In this paper we have investigated three different ensembles to analyze 
the variation of particle yields with the 
multiplicity of charged particles produced in proton-proton collisions at the center-of-mass energy
of $\sqrt{s}$~=~7~TeV. 
It is  interesting to note that all three ensembles lead to the same results when the 
multiplicity of charged particles $\dNchdeta$ exceeds about 20. This could be interpreted as reaching the thermodynamic limit
since the three  ensembles lead to the same results.  The total number of hadrons in the final state is
 of the order of 300 for the highest  multiplicity class when integrated over the full rapidity interval.
Another observation is that the density tends to a constant with increasing multiplicity.
It would be of interest to extend this analysis to higher beam energies and higher multiplicity intervals.\\[1cm]
\large{{\bf Acknowledgments}}\\
\normalsize
One of us (J.C.) gratefully thanks  the National Research Foundation of South Africa for financial support.
N.S. acknowledges the support of SERB Ramanujan Fellowship
(D.O. No. SB/S2/RJN- 084/2015) of the Department of Science and Technology of India.
B.H. acknowledges the support of the Universit\'e de Strasbourg Institute for Advanced Study.

\bibliographystyle{epjc}
\bibliography{sch_pp}
\end{document}